\documentstyle[12pt]{article}
\begin{document}
\newcommand{\be}{\begin{equation}}
\newcommand{\ee}{\end{equation}}
\begin{center}

{\bf Soft and hard processes in QCD}\\

\vspace{2mm}

I.M. Dremin\footnote{Email: dremin@lpi.ru}\\
\vspace{2mm}

Lebedev Physical Institute, Moscow 119991\\

\end{center}

\begin{abstract}
QCD equations for the generating functions are applied to separate soft and
hard jets in $e^+e^-$-processes of multiparticle production. The dependence 
of average multiplicities and higher moments of multiplicity distributions 
of particles created in a ``newly born" soft subjets on the share of energy 
devoted to them is calculated in fixed coupling gluodynamics. This dependence
is the same as for the total multiplicity up to a constant factor if soft jets
are defined as those carrying out a fixed share of initial energy at all 
energies. The constant factor depends on this share in a non-trivial way.
Other definitions are also proposed.
The relation between these quantities for soft and hard processes is discussed.
\end{abstract}

Key words: gluon, jet, multiplicity\\

In multiparticle production, it is quite common procedure to separate all
processes into the soft and hard ones. Even though the intuitive approach is 
appealing, the criteria of the separation differ. It is shown below that QCD 
equations for the generating functions can be applied to this problem. It is
demonstrated how the average multiplicities of soft and hard processes depend
on the parameter which is used to distinguish them. The same method can be
applied to any moment of the multiplicity distributions as is explicitely 
shown for the second moment (dispersion).

The QCD equations for the generating functions (functionals) are known since 
long ago (e.g., see the book \cite{dkmt}). This is the system of two 
integro-differential equations which describe the quark and gluon jets
evolution. They are quite useful for prediction and description of many 
properties of high energy jets (for the reviews see, e.g., 
\cite{dre1, dgar, koch}). It has been found that the main qualitative features
of the process can be safely predicted by considering the single equation for 
gluon jets evolution. In that way one neglects quarks and treats the 
gluodynamics in place of the chromodynamics. Moreover, its solution may be
further simplified if one disregards the running property of the QCD coupling 
strength and considers it as a fixed one (see papers \cite{dhwa}). To avoid
some technicalities, we adopt this approach in what follows and treat the 
multiplicity distributions of gluon jets. Both quark and gluon jets with 
running coupling strength will be considered in the QCD context elsewhere.

When an initial gluon splits into two gluons (subjets), its energy $E$ is 
additively shared among them, and the multiplicity of the whole process is 
a sum of multiplicities of these two subjets. The energy dependence of mean
multiplicity of particles created in a subjet, which carries out some share
of initial energy $xE$ with a fixed value of $x$, must be the same as for the 
initial jet if gluons are equivalent. In
experiment it is more convenient to deal with values of $x$ ranging in some
finite interval to get enough statistics. One of the ways is to separate
all subjets into soft and hard ones if the parameter $x$ is smaller or larger
than some $x_0$. We show how properties of these two sets behave with energy.
We shall also consider the case when the parameter $x_0$ depends on
initial energy.

If the probability to create $n$ particles\footnote{In what 
follows, we adopt the local parton-hadron duality hypothesis with 
no difference between the notions of particles and partons up 
to some irrelevant factor.} in a jet is 
denoted as $P_n$, the generating function $G$ is defined as
\be
G(z,y) = \sum_{n=0}^{\infty }P_n(y)(1+z)^{n},                    \label{3}
\ee
where $z$ is an auxiliary variable,
$y=\ln (p\Theta /Q_0 )=\ln (2Q/Q_{0})$ is the evolution parameter, defining
the energy scale, $p$ is the initial momentum, $\Theta $ is the angle of 
the divergence of the jet (jet opening angle), assumed here to be 
fixed, $Q$ is the jet virtuality,  $Q_{0}=$ const.

The gluodynamics equation for the generating function is written as
\be
dG/dy= \int_{0}^{1}dxK(x)\gamma _{0}^{2}[G(y+\ln x)G(y+\ln (1-x)) - 
G(y)],      \label{geq}
\ee
where
\begin{equation}
\gamma _{0}^{2} =\frac {6\alpha _S}{\pi } ,                \label{52}
\end{equation}
$\alpha _{S}$ is the coupling strength and the kernel $K(x)$ is
\begin{equation}
K(x) = \frac {1}{x} - (1-x)[2-x(1-x)] .    \label{53}
\end{equation}
One should not be surprised that the shares of energy $x$ and $1-x$ devoted to 
two gluons after the initial one splits to them enter asymmetrically in this 
equation. Surely, the initial equation is fully symmetrical. The asymmetry is 
introduced when the phase space is separated in two equally contributing parts
and one of the jets with the share $x$ is called as a ``newly born" one (for
more details see \cite{dkmt}). Therefore we shall call soft processes those
where soft newly born jets are produced, i.e. those where $x$ is small enough
($x\leq x_0\ll 1$). In $e^+e^-$-experiments, this would correspond to 
considering soft newly born gluon jets with energies $E_g\leq x_0E\ll E$ 
in 3-jet events.

Before separating soft and hard jets, let us stress that, at a given energy, 
this is an 
additive procedure for probabilities $P_n=P_{ns}+P_{nh}$ and, consequently,
for $G=G_s+G_h$, where indices $s$ and $h$ are for soft and hard processes, 
correspondingly. It is convenient to rewrite the generating function in terms 
of unnormalized factorial moments
\be
{\cal F}_{q} = \sum_{n} P_nn(n-1)...(n-q+1) =
 \frac {d^{q}G(z)}{dz^{q}}\vline _{z=0},         \label{calf}
\ee
so that
\be
G=\sum_{q=0}^{\infty }\frac{z^q}{q!}{\cal F}_{q}.            \label{gf}
\ee
The low rank moments are
\be
{\cal F}_{1}=\langle n\rangle, \;\;\; {\cal F}_{2}=\langle n(n-1)\rangle=D^2+
\langle n\rangle ^2-\langle n\rangle         \label{cal1}
\ee
and $D$ is the dispersion
\be
D^2= \langle n^2\rangle -\langle n\rangle ^2.     \label{disp}
\ee
It is seen from eq. (\ref{calf}) that unnormalized moments are additive also.
To retain the additivity property we define the normalized factorial moments
for soft and hard jets with the normalization to the total mean multiplicity
but not to their multiplicities
\be
F_q=\frac{{\cal F}_q}{\langle n\rangle ^q}=\frac{{\cal F}_{qs}+{\cal F}_{qh}}
{\langle n\rangle ^q}=F_{qs}+F_{qh}.     \label{fqn}
\ee
Thus the total multiplicity is in the denominator. The additivity would be lost
if soft and hard values are normalized to their average multiplicities 
$\langle n_s\rangle $ and $\langle n_h\rangle $. Introducing $f_q=F_q/q!$ we
write
\be
G=\sum_{q=0}^{\infty }z^q\langle n\rangle ^qf_q.   \label{gfq}
\ee
The scaling property of the fixed coupling QCD \cite{dhwa} allows to look for 
the solution of the equation (\ref{geq}) with 
\be
\langle n\rangle \propto \exp (\gamma y) \;\;\;\; (\gamma = const)  \label{ny}
\ee
and get the system of iterative equations for $f_q$
\be
\gamma qf_q=\gamma _0^2\int_0^1dxK(x)[(x^{\gamma q}+(1-x)^{\gamma q}-1)f_q+
\sum_{m=1}^{q-1}x^{\gamma m}(1-x)^{\gamma (q-m)}f_mf_{q-m}].   \label{eqfq}
\ee
By definition $f_1=1$ and one gets at $q=1$ the relation between $\gamma $
and $\gamma _0$
\be
\gamma =\gamma _0^2\int_0^1dxK(x)(x^{\gamma }+(1-x)^{\gamma }-1)
=\gamma _0^2M_1(1,\gamma ),                                   \label{gam}
\ee
where
\be
M_1(z,\gamma )=\int_0^zdxK(x)(x^{\gamma }+(1-x)^{\gamma }-1). \label{m_1}
\ee
Let us point out that eq. (\ref{gam}) is derived from the equation for mean 
multiplicities which follows from eq. (\ref{geq}):
\be
\langle n(y)\rangle ^{'} =\int_0^1dx\gamma _{0}^{2}K(x)(\langle n(y+\ln x)\rangle 
+\langle n(y+\ln (1-x))\rangle -\langle n(y)\rangle ).  \label{nav}
\ee
These results are well known \cite{dhwa}. Here, we would like to consider 
eq. (\ref{gam}) in more detail. As follows from eq. (\ref{nav}), the first 
two terms in the brackets correspond to mean multiplicities of two subjets,
and their sum is larger than the third term denoting the mean multiplicity
of the initial jet (all divided by $E^{\gamma }$). Therefore, the integrand 
is positive, and eq. (\ref{gam}) defines the anomalous dimension $\gamma $.
This does not contradict to the statement that for a given event the total 
multiplicity is a sum of multiplicities in the two subjets because the 
averages in eq. (\ref{nav}) are done at different energies. 

For small enough $\gamma $ and $\gamma _0$ one gets
\be
\gamma \approx \gamma _0(1-0.458\gamma _0+ 0.213\gamma _0^2).   \label{gam0}
\ee
For the second moment one gets from eq. (\ref{eqfq}) at $q=2$ for small 
$\gamma $
\be
F_2 \approx \frac{4}{3}(1-0.31\gamma ).     \label{f2g}
\ee
Now, according to the above discussion we define soft jets as those with 
sum of energies of belonging to them particles less than some $x_0E$. First, 
consider $x_0$=const and small. Then 
we should choose the upper limit of integration in eq. (\ref{eqfq}) equal to
$x_0$. Therefore the moments of soft processes $f_{qs}$ are calculated as
\be
\gamma qf_{qs}=\gamma _0^2\int_0^{x_0}dxK(x)[(x^{\gamma q}+(1-x)^{\gamma q}-1)f_q+
\sum_{m=1}^{q-1}x^{\gamma m}(1-x)^{\gamma (q-m)}f_mf_{q-m}].     \label{eqfs}
\ee
One should not be confused that the total moments (obtained from the 
average multiplicities of both soft and hard jets) are in the integrand of eq.
(\ref{eqfs}). This is related to the difference between the notions of 
multiplicity in a given event and their averages discussed above. The 
integration over small $x$ up to $x_0$ chooses just the mean multiplicity of 
particles belonging to soft jets $\langle n_s\rangle $ while the integration 
from $x_0$ to 1 gives that for hard jets.

For $q=1$ one gets from (\ref{eqfs})
\be
\frac{\langle n_s\rangle}{\langle n\rangle}=
\frac{M_1(x_0,\gamma)}{M_1(1,\gamma )}.            \label{nii}
\ee
For small $x_0$ it is
\be
\frac{\langle n_s\rangle}{\langle n\rangle} \approx  \frac{\gamma_0^2}{\gamma ^2}
x_0^{\gamma }N_1(x_0, \gamma ),        \label{nsn}
\ee
\be
N_1(x_0, \gamma )=1-\gamma ^2x_0^{1-\gamma }-\frac{2\gamma }{1+\gamma }x_0+
\frac{\gamma ^2(3+\gamma )}{4}x_0^{2-\gamma }+\frac{3\gamma }{2+\gamma }x_0^2 -
\frac{\gamma ^2 (2+\gamma )}{3}x_0^{3-\gamma }.     \label{n1g}
\ee
Thus we have found the energy dependence of mean multiplicity of particles in 
a set of subjets with low energies $E_s\leq x_0E$. As expected for constant
$x_0$, it is the same as the energy dependence of the total multiplicity with 
a different factor in front of it. Namely this dependence should be 
checked first in experimental data. Imposed on one another, these figures 
should coincide up to a normalization factor (\ref{nii}). This would confirm 
universality of gluons in jets.
 
Quite interesting is the non-trivial dependence of the normalization factor 
in eq. (\ref{nii}) on the parameter $x_0$, which does not coincide simply with
$x_0^{\gamma }$. It reflects the structure of QCD kernel $K(x)$. The main 
dependence on the cut-off parameter $x_0$ is given for $x_0\ll 1$ by the 
factor $x_0^{\gamma }$ with the same power as in dependence of
total multiplicity on energy. This corresponds 
to subjets with the largest energy of the set. However, with increase of $x_0$, 
this dependence is modified according to eqs (\ref{nii})-(\ref{n1g}). 
The negative corrections become more important in eq. (\ref{n1g}). They are
induced by subjets with energies lower than $x_0E$. The decrease of the
normalization factor corresponds to diminishing role of very low energy
jets at higher initial energies. This should be also checked in experiment.

If plotted as a function of the maximum energy in a set of jets $\epsilon_m$, 
the mean multiplicity is
\be
\langle n_s\rangle \propto \epsilon _m^{\gamma }[1-
\gamma ^2\left (\frac {\epsilon _m}{E}\right )^{1-\gamma }-
\frac{2\gamma }{1+\gamma }\left (\frac {\epsilon _m}{E}\right )+
\frac{\gamma ^2(3+\gamma )}{4}\left (\frac {\epsilon _m}{E}\right )^{2-\gamma }+
\frac{3\gamma }{2+\gamma }\left (\frac {\epsilon _m}{E}\right )^2 -
\frac{\gamma ^2 (2+\gamma )}{3}\left (\frac {\epsilon _m}{E}\right )^{3-\gamma }].
\label{eps}
\ee
It reminds eq. (\ref{ny}) with the correction factor in the brackets.

This is the consequence of the scaling property of the fixed coupling QCD which
results in the jets selfsimilarity. The relative weights of soft and hard 
processes are determined by the factor $\gamma_0^2/\gamma ^2$ as seen from eqs
(\ref{nsn}), (\ref{n1g}). They can be used to find out this ratio in experiment.

For $q=2$ we obtain
\be
F_{2s}=[0.5F_2M_1(x_0, 2\gamma )+M_2(x_0,\gamma )]/M_1(1,\gamma ),  \label{fii}
\ee
where
\be
M_2(z,\gamma )=\int_0^zdxK(x)x^{\gamma }(1-x)^{\gamma}.   \label{i2g}
\ee
For small $x_0$ one gets
\be
F_{2s}=\frac{\gamma _0^2}{\gamma ^2}x_0^{\gamma }[0.25F_2x_0^{\gamma }
N_1(x_0, 2\gamma )+2N_2(x_0, \gamma )],      \label{f2n}
\ee
\be
N_2(x_0, \gamma )=1-\gamma \frac{2+\gamma }{1+\gamma }x_0+\gamma \frac {6+3\gamma 
+\gamma ^2}{2(2+\gamma )}x_0^2.              \label{n2g}
\ee
Again, the main dependence on the cut-off parameter $x_0$ is provided 
by the factor $x_0^{\gamma }$. 

\begin{table}[b]
 \caption{
The values of mean multiplicity and second normalized factorial moment
for different values of the coupling strength and cut-off parameters $x_0$
}
 \begin{center}
\begin{tabular}{|c||c|c||c|c||c|c|} \hline
     & \multicolumn{2}{c||}{$\gamma=0.5,$ $\gamma_0=0.7$}&      
\multicolumn{2}{c||}{$\gamma=0.4,$ $\gamma_0=0.516$}&
\multicolumn{2}{c|}{$\gamma=0.3,$ $\gamma_0=0.36$}     \\ \hline
$x_0$& $n_s/n$      & $F_{2s}$   & $n_s/n$    &  $F_{2s}$      & $n_s/n$  &  $F_{2s}$   \\   \hline
$0.1$& 0.543        & 0.39       & 0.605      & 0.66         &  0.680    &   0.76  \\
$0.2$& 0.702        & 0.51       & 0.746      & 0.83         &  0.798   &   0.91   \\
$0.3$& 0.798        & 0.59       & 0.829      & 0.93         &  0.865   &   1.00  \\   \hline
\end{tabular}                    \\
 \end{center}
\end{table}

Using these equations we have calculated the mean multiplicities and second 
moments of multiplicity distributions for soft jets. They are shown in Table 1 
for different choices of
$\gamma $ and $\gamma _0$ considered as the most realistic ones in previous 
studies. Note that $M_1(z,2\gamma )=0$ for $\gamma =0.5$ (and so is 
$N_1(x_0, 2\gamma )$). One can notice that at larger $x_0$ the values in Table 1
decline from $x_0^{\gamma }$-behaviour in accordance with eqs (\ref{nii}),
(\ref{fii}).

The values for hard jets are obtained by subtracting these results from 
values for the total process. Small values of the second factorial moments 
do not imply that multiplicity distributions in soft jets are sub-poissonian 
because they are normalized 
to the total mean multiplicity. To get the genuine second factorial moments
for these processes one should divide the numbers in $F_{2s}$-columns to
squared values in $n_s/n$-columns. In this way one gets quite large numbers
so that these processes are super-poissonian but note that the genuine moments 
are not additive anymore. However, the statement about the widths of the
distributions can be confronted to experimental data as well.

In principle, other definitions of soft jets are possible with $x_0=x_0(E)$. 
Then one should solve the equation 
\be
\frac{d\langle n_s\rangle }{dE} =E^{\gamma -1}
\gamma _{0}^{2}M_1(x_0(E), \gamma ),  \label{navs}
\ee
which follows from eq. (\ref{nav}). For example, one can choose the jets 
with energies less than some fixed constant independent of the initial 
energy. This would imply $\epsilon _m=$const or $x_0(E)\propto 1/E$, and the exact integration of 
eq. (\ref{navs}) is necessary. However, for qualitative estimates, eqs 
(\ref{nsn})-(\ref{eps}) can be used. They show that the average multiplicity 
tends to a constant at high energies corresponding to the multiplicity at 
the upper limit. At lower energies, it slightly increases with energy due to 
increasing role of jets with energies closest to their upper limit.

It is well known that for running coupling the power dependence $s^{\gamma /2}$
is replaced by $\exp(c\sqrt {\ln s})$. The qualitative statement about the 
similar energy behaviour of mean multiplicities in soft and inclusive processes 
should be valid also.

The above results can be confronted to experimental data if soft jets are 
separated in 3-jet events. However, in our treatment we did not consider the 
common experimental cut-off which must be also taken into account.
This is the low-energy cut-off imposed on a soft jet for the third jet to be
observable. It requires the soft jet not to be extremely soft. Otherwise the
third jet is not separated and the whole event is considered as a 2-jet one.
Thus the share of energy must be larger than some $x_1$, and the integration 
in eq. (\ref{m_1}) should be from $x_1$ to $x_0$. For $x_1\leq x_0\ll 1$ one
gets
\be
\frac{\langle n_s\rangle}{\langle n\rangle}=
\frac{\gamma_0^2}{\gamma ^2}[v(x_0)-v(x_1)],       \label{vx}
\ee
where the function $v(x)$ is easily guessed from eqs (\ref{nsn}), (\ref{n1g}).
At $x_1\ll x_0\ll 1$ eq. (\ref{nsn}) is restored. 

The cumulant moments of the distribution are not additive because they are 
obtained as derivatives of the generating function logarithm which 
is not additive for additive $G$.
Thus $H_q$-moments \cite{13, dnec} are not additive also. Nevertheless,
the role of hard jets can be traced by the preasymptotical oscillations 
of $H_q$. These oscillations are induced by the terms of the kernel $K$ 
additive to the $1/x$-term. Thus the oscillations of $H_q$ are the sensitive 
test of the shape of non-infrared terms in the QCD kernels and their integral
contributions. Namely these terms contribute much to hard processes
because they favour larger values of $x$. The stronger is their influence,
the closer to zero should be the intercept of $H_q$ with the abscissa axis.
It would be interesting to get experimental information about the behaviour
of $H_q$ for soft and hard jets separately.

In conclusion, the separation of soft and hard jets according to the
share of energy devoted to the ``newly born" jet is proposed. If this is 
done, the experimentally measured values of mean multiplicities and other 
multiplicity distribution parameters of particles belonging to the soft jet 
can be compared with the obtained above theoretical predictions at different 
values of this share of energy. For a constant share, this dependence is
the same as for the average total multiplicity but with non-trivial 
$x_0$-dependence of the factor in front of it. Some predictions are obtained for 
energy dependent cut-offs. The conclusions can be confronted to experiment.  \\

{\bf Acknowledgments}\\

I am grateful to V.A. Nechitailo and E. Sarkisyan-Grinbaum for useful comments.
This work has been supported in part by the RFBR grants 03-02-16134, 
04-02-16445-a, NSH-1936.2003.2.\\

\end{document}